\documentclass[fleqn,12pt]{article}
\usepackage{amsmath}
\usepackage{amsfonts,amssymb,latexsym}

\newcommand{\ft}[2]{{\textstyle\frac{#1}{#2}}}
\def\tilde{\widetilde}
\def\1bar{1\hskip -.275cm -}
\def\2bar{2\hskip -.275cm -}
\def\3bar{3\hskip -.275cm -}

\newsavebox{\uuunit}

\newcommand{\pls}{\!+\!}

\newcommand{\mis}{\!-\!}

\def\a{\alpha}
\def\b{\beta}
\def\g{\gamma}
\def\l{\lambda}
\def\d{\delta}
\def\e{\epsilon}
\def\t{\theta}

\def\s{\sigma}

\def\G{\Gamma}

\def\p{\partial}

\newcommand{\be}{\begin{equation}} \newcommand{\ee}{\end{equation}}
\newcommand{\bea}{\begin{eqnarray}} \newcommand{\eea}{\end{eqnarray}}
\newcommand{\ben}{\begin{displaymath}}
\newcommand{\een}{\end{displaymath}}
 
\newcommand{\nn}{\nonumber}

\makeatletter
\let\old@makecaption=\@makecaption
\def\@makecaption{\small\old@makecaption}
\makeatother

\makeatletter \@addtoreset{equation}{section} \makeatother

\makeatletter
\let\old@startsection=\@startsection
\renewcommand{\@startsection}[6]{\old@startsection{#1}{#2}{#3}{#4}{#5}{#6\mathversion{bold}}}
\makeatother

\newcommand{\hypref}[2]{\ifx\href\asklfhas #2\else\href{#1}{#2}\fi}

\newenvironment{myeqnarray}{\arraycolsep0pt\begin{eqnarray}}{\end{eqnarray}\ignorespacesafterend}
\newenvironment{myeqnarray*}{\arraycolsep0pt\begin{eqnarray*}}{\end{eqnarray*}\ignorespacesafterend}

\def\[{\begin{equation}}
\def\]{\end{equation}}
\def\<{\begin{myeqnarray}}
\def\>{\end{myeqnarray}}

\def\sqr#1#2{{\vcenter{\vbox{\hrule height.#2pt
 \hbox{\vrule width.#2pt height#1pt \kern#1pt
 \vrule width.#2pt}\hrule height.#2pt}}}}

\def\square{\mathop{\mathchoice{\sqr{12}{15}}{\sqr{9}{12}}{\sqr{6.3}{9}}{\sqr{4.5}{9}}}}

\newcommand\da{\dot  \alpha}
\newcommand\db{\dot \beta}
\newcommand\dg{\dot \gamma}

\ifx\href\asklfhas\newcommand{\href}[2]{#2}\fi

\newdimen\squaresize \squaresize=12pt
\newdimen\thickness \thickness=0.7pt

\def\square#1{\hbox{\vrule width \thickness
   \vbox to \squaresize{\hrule height \thickness\vss
      \hbox to \squaresize{\hss#1\hss}
   \vss\hrule height\thickness}
\unskip\vrule width \thickness} \kern-\thickness}

\def\cut#1{\hbox{\vrule width-1 \thickness
   \vbox to \squaresize{\hrule height-1 \thickness\vss
      \hbox to \squaresize{\hss#1\hss}
   \vss\hrule height-1\thickness}
\unskip\vrule width +4 \thickness} \kern-\thickness}

\def\vsquare#1{\vbox{\square{$#1$}}\kern-\thickness}

\begin{document}
\global\parskip=4pt

\rightline{DISTA-UPO-05}
\rightline{CERN-PH-TH/2005-207}
\rightline{25/10/2005}
\vskip 1.5 truecm

\LARGE
\bf

\begin{center}
{\Huge Partition Functions of Pure Spinors}
\end{center}

\normalsize
\rm

\vskip 1.5 truecm
\normalsize
\centerline{\large P.~A.~Grassi$^{~a,b,c,~}$\footnote{~pgrassi@cern.ch},
 J.~F. ~Morales Morera$^{c,~}$\footnote{fmorales@cern.ch}}
 \vskip .7 truecm
\centerline{$^{(a)}$
{\it Centro Studi e Ricerche E. Fermi,} }
\centerline{\it 
Compendio Viminale, I-00184, Roma, Italy,}
\vskip .3cm
\centerline{$^{(b)}$ {\it DISTA, Universit\`a del Piemonte Orientale}}
\centerline{\it Via Bellini 25/g,  Alessandria, 15100, Italy,}
\centerline{{\it and INFN, sezione di Torino,}}
\vskip .3cm
\centerline{$^{(c)}$
{\it
CERN, Theory Unit, CH-1211 Geneve, 23, Switzerland}}

\vskip 0.7 truecm
\normalsize
\bf
\centerline{Abstract}

\rm
\begin{quotation}
We compute partition functions describing multiplicities and
charges of massless and first massive string states of
pure-spinor superstrings in $3,4,6,10$ dimensions. At the massless
level we find a spin-one gauge multiplet of minimal supersymmetry
in $d$ dimensions. At the first massive string level we find a massive
spin-two multiplet. The result is confirmed by a direct analysis of
the BRST cohomology at ghost number one. 
The central charges of the pure
spinor systems are derived in a manifestly $SO(d)$ covariant way
confirming that the resulting string theories are critical.
A critical string model with ${\cal N}=(2,0)$ supersymmetry
in $d=2$ is also described.
\end{quotation}

\newpage

\tableofcontents

\section{Introduction}

During the last years, one of the fundamental piece of work in
string theory is the construction of a quantizable model of
superstring in 10 dimensions with manifest super-Poincar\'e
invariance \cite{Berkovits:2000fe}. The progress in understanding
gauge/gravity
correspondences \cite{Maldacena:1997re} makes clear
than an understanding of string backgrounds in presence of RR
fluxes are crucial to go beyond the supergravity level. The
super-Poincar\'e invariant formulation \cite{Berkovits:2000fe} of ten
dimensional superstring treats NSNS (Nevue-Schwarz) and RR(Ramond) fields on the same
footing and it is the natural candidate to address this question (see \cite{Berkovits:2002zk} for a review).

The covariant formulation \cite{Berkovits:2000fe} is based on a set of worldsheet
bosonic fields $\l^{\a}$ (and its complex conjugate $w_\a$)
transforming as an $SO(10)$ spinor and satifsying a pure spinor
constraint $\l \gamma^m \l=0$. The pure spinor system is tensored with
free fermions $(\theta^\a,p_\a)$ and bosons $x^m$ in such a way that
the total conformal charge is zero and the Lorentz generators associated to the
spinorial variables (GS variables and ghost fields) have the same
double poles of the Lorentz generator for worldsheet fermions of
RNS. Additionally, the physical spectrum is constructed on a
free-field action (for flat space) using a single non-hermitian
BRST charge which is nilpotent when the pure spinor constaint is satisfied.
 It has been shown that
the string spectrum identified with the BRST cohomology coincides
with the light-cone spectrum of the GS strings
\cite{Berkovits:2000nn,Berkovits:2004tw,Aisaka:2004ga}. This spectrum is found by solving
the pure spinor constraint in an $SO(8)$ covariant way.
 Several checks have been performed for
tree level and one-loop amplitudes
\cite{Berkovits:2000ph,Berkovits:2004px,Anguelova:2004pg}.

It is important to recall that even if the pure-spinor
formalism is manifestly super-Poincar\'e invariant, covariant
computations are rather cumbersome due to the ghost contraints
which have to be implemented at any stage. In papers
\cite{Grassi:2001ug, Grassi:2002tz, Grassi:2002xv} the authors
reproduce the massless spectrum of open and closed superstrings
using some additional ghost fields and imposing a grading constraint
on the functional space to retrieve the correct constraints.
Several studies \cite{Aisaka:2002sd, Aisaka:2003mw,
Oda:2001zm,Matone:2002ft,Chesterman:2002ey} followed the original
papers extending the analysis in different directions.

Recently,  in \cite{Berkovits:2005hy},
the authors exploited localization techniques
to compute the zero mode part of the partition function
for pure spinors in $d=10$. The results were written
in a $SO(10)$-covariant form
and linked directly to the super-Yang-Mills multiplet dynamics describing
the massless modes of the
open ten-dimensional superstring.
Indeed, it has been shown that not only the physical states (8 states of the
vector multiplet and 8 states of the gluino) are represented, but also the
complete set of ghosts and the antifields implementing the equations of motion.
What about massive string modes? Massive string states enter in a
somehow trivial way in most of the  pure spinor string amplitudes studied so far.
 Even the simplest open string amplitude describing an open string in presence
 of a constant magnetic field has not been yet reproduced inside this formalism
 (see \cite{Schiappa:2005mk} for a study of the massless part).
 The main missing ingredient is a covariant description of
 the spectrum of the theory. Aim of this paper is to considering the first massive
 string state, as a first step towards a $SO(10)$ covariant string partition
 function. This is a crucial ingredient for the study of superstring
 spectra in non-trivial backgrounds like $AdS_{5} \times S^{5}$.
In particular, it would be nice to explain, by a direct computation
in the pure spinor formalism, the spectrum and mass formula
for string states on $AdS_5\times S^5$ found via KK analysis
in \cite{Bianchi:2003wx}-\cite{Beisert:2004di}
 (see \cite{Morales:2004xc} for generalizations  to Dp-brane geometries).

In this paper we consider string theories based on pure
spinors in dimensions $d=4,6,10$\footnote{We advise the reader that our
definitions of pure spinors in $d=4,6$ differ from those used in
\cite{Berkovits:2005hy} in the spinor representation chosen for $\l^\a$.
This explains why partition functions and central charges differ from those
in that reference.}.
We refer the reader to
\cite{Grassi:2005sb,Wyllard:2005fh}
for details in the definitions of the pure spinor strings
in $d\neq 10$.\footnote{Related work on the N=2 string formulation appeared
in \cite{Berkovits:2005bt,Chandia:2005fi}. Related work on the massless
cohomology of lower dimensional models appeared also in the interesting paper by A. Movshev
\cite{Movshev:2005cq}. }
We will compute the massless and first
massive string state partition functions. As in \cite{Berkovits:2005hy}, we count
states in a $SO(d)$ covariant way without solving the pure spinor constraint.
From the massless partition
function we extract the central charge of the pure spinor system and show that
the resulting superstrings are critical in any dimension.
In addition we will show that the massless partition functions of the
pure spinor open string have precisely the degrees of freedom to describe
${\cal N}=1$ SYM in $d=4,6,10$.  Closed superstrings based on these pure spinors
realize ${\cal N}=2$ supergravities in $d=4,6,10$ with RR fields and branes
and are suitable for studies of holography.
 Superstrings on $AdS_5\times S^1$ has been recently proposed \cite{Klebanov:2004ya}
as holograhic duals of ${\cal N}=1$ gauge theories in $d=4$. The existence of
critical pure-spinor superstrings realizing minimal Yang-Mill theories in any dimension open a new handle to covariant quantization of these AdS strings.

We will follow the strategy in \cite{Berkovits:2005hy}. Rather
 than compute the cohomology of the BRST operator defining the physical
 spectrum, we compute the $(-)^F$ weighted string partition function,
 $F$ being the worldsheet fermionic number keeping track also of
 $SO(d)$ charges. Since BRST operator is odd
 under $(-)^F$, paired states do not contribute to this index and therefore the
 string partition function counts only states in the cohomology. The string
 partition function results provide us then with shortcuts to the
 often lengthly cohomology  analysis.

The paper is organized as follows: In Section \ref{secpure} we introduce
the pure spinor models in $d=4,6,10$ and compute their
string partition functions. In Section \ref{seccohomology} we compute
the cohomology at ghost number one. We find perfect agreement with the results
coming from the string partition function. In section \ref{seclow} we introduce a
pure spinor system with ${\cal N}=(2,0)$ supersymmetry in $d=2$
and one with minimal supersymmetry in $d=3$. In each case we derive the
massless and first massive string spectrum both from counting of pure spinor
states and cohomology analysis. In Section \ref{secconclusions} we summarize
our results and comment on future directions. Finally in appendix \ref{secmult}
we collect some details on the construction of massive multiplets in $d=4,6,10$, 
and in appendix B we give details of anomalous Ward identities. 

\section{Pure spinors}
\label{secpure}

 Superstrings based on pure spinors are defined be the sigma model
\be S= {1\over 2\pi \a'} \int d^{2}\sigma
\Big(\p x^\mu \bar\p x^\mu+ p_{A} \bar\p \t^{A} +
w_{A} \bar\p \l^{A}\Big)
\label{action} \ee Here $x^\mu$ describes the
string coordinates in $\mathbb{R}^{d}$, $\theta^{A}$ are
anticommuting variables and $p_{A }$ their conjugate momenta.
The field
$\lambda^{A}$ satisfy the pure spinor constraint \be
 \lambda \gamma^\mu \lambda=0\label{pure}
\ee
 The pure spinor constraint (\ref{pure}) induces a gauge invariance on $w_A$
(the conjugate momentum of $\lambda^A$):
\be
 \delta w_A = \Lambda_{\mu} (\g^{\mu} \l)_{A}
\label{inv}
\ee
Indices $\mu$, $A$(up) and $A$(down) run over the vector ${\bf V_d}$
and spinor representations ${\bf S_d}$, ${\bf \bar{S}_d}$ respectively of
the Lorentz group $SO(d)$. Spinors are chosen to be
 Dirac in $d=4$ , symplectic
Majorana Weyl in $d=6$ and Majorana-Weyl in $d=10$. The choice of
the spinor representations ${\bf S_d}$
is such that the vector representation ${\bf V_d}$ always appear in the
product ${\bf S_d}\times {\bf S_d}$ of two $\l$'s.
 The dimensions $S_d$ and Lorentz content in the various dimensions
are resumed in table 1.
\begin{table}[h]
\begin{center}
\begin{tabular}{|l|l|l|l|}
\hline
      d & Lorentz+R &  Dynkin Labels & $  S_d$ \\ \hline
      4 & $SU(2)\times SU(2)$ &  $[\ft{1}{2},0]+ [0,\ft{1}{2}]$& 4 \\
   6 & $SO(6)\times SU(2)$ &  $[001]_{\ft{1}{2}}$&  8 \\
  10 & $SO(10)$ &  $[00001]$ & 16\\
 \hline
     \end{tabular}
\label{lambda1}
\caption{Spinors $\lambda^\alpha,\theta^\a$ in $d$ dimensions.}
 \end{center}
 \end{table}
 The resulting string theories are critical in any dimension. This
 can be seen by notice that the naive central charge  of the pure spinor
 system $c_{\l,w}=2 S_d-d$ cancels against that
of the free $x^m,\theta$ system $c_{x,p,\theta}=-2 S_d + d$. This
naive counting of degrees of freedom will be confirmed below by a
$SO(d)$ covariant derivation of the central charges from the pure
spinor partition function.

We will organize the states according to the $U(1)$ charge $\Delta$:
\be
 \Delta=n_\lambda+n_\t+2\,n_x+3\, n_w+3\, n_p
\ee
which is clearly a symmetry of (\ref{action}) if we assign to $\alpha '$
charge $\Delta=4$.
 The spectrum of string states and $\Delta$-charges can be read from the
character valued partition function
\be
 Z(q|t)={\rm tr}_{\cal H} \, (-)^F\, q^{L_0} \, t^\Delta
\label{zpart}
\ee
with $L_0$ the string level and $F$ the fermion number. The trace runs
over the space of polynomials of the string modes
$\l,\t,p,w,x$ (and their worldsheet
derivatives) satisfying the pure spinor constraint
(\ref{pure}) and invariant under (\ref{inv}). The Fock space can then be
written as a polynomials in the string modes:
 \be
{\cal H}=\{ F(\l_n, \t_n,
x_n, p_{n>0}, w_{n>0})|0\rangle~~ |~~
\sum_{m}\lambda_{n-m} \gamma^\mu \l_{m}=0~;~~\delta F=0 \}
\label{fock}
\ee
 with $m,n=0,1,..$.  For simplicity we
will always omit the contribution of bosonic zero
modes $x_0$, i.e. we focus on the zero momentum spectrum.

The results will be written as
 polynomials in Lorentz representations made out of products of
 vector ${\bf V_d}$ and spinor representations ${\bf S_d}$, ${\bf \bar{S}_d}$.
 In particular the contribution of a free worldsheet
 field $\Phi_n$ in a given representation ${\cal R}$ with scaling $t^a$
 will be written as:
 \bea
 {1\over (1-t^a\, q^n)^{\bf R}} &=& 1+t^a\, q^n\, {\bf R}+t^{2a}\, q^{2n}\,
 ({\bf R}\times {\bf R})_{\rm sym}+\ldots\nn\\
(1-t^a\, q^n)^{\bf R} &=& 1-t^a\, q^n\, {\bf R}+t^{2a}\, q^{2n}\,
 ({\bf R}\times {\bf R})_{\rm antisym}+\ldots
 \eea
  for a bosonic/fermionic worldsheet mode $\Phi_n$
respectively. The blind partition function follows from replacing the
representation ${\bf R}$ by its dimension.

Before entering the details of the computation let us
sketch our general strategy to count pure spinor states.
 Written in string modes, the pure spinor constraint reads
\bea
&&  \l_0 \gamma^m \l_0=0 \nn\\
&& \l_0 \gamma^m \l_1=0
\eea
and so on. This condition restricts the number of representations that
appear in the tensor product of two pure spinors.
 Explicitly:
 \bea
\l_0\times \l_0:&& ({\bf S_d}\times {\bf S_d})_{\rm sym}-{\bf V_d}\nn\\
\l_0\, \times \l_1: && ({\bf S_d}\times {\bf S_d})-{\bf V_d}
\label{prod1}
\eea
and so on. In other words, polynomials in $\l$ are given
by symmetric product of the spinor representation ${\bf S_d}$
where the vector representation ${\bf V_d}$
in each bispinor product is deleted.

Analogously gauge invariance requires that $F$
depends on $w$ only via the gauge invariant
combinations  $J_A^B={\cal P}_{BD}^{AC} w_C \l^D$
with ${\cal P}_{BD}^{AC}$ a projector on the gauge invariant components:
\bea
d=4:  && [1,0]+[0,1]+2[0,0]\nn\\
d=6: && [011]_0+[000]_1+[000]_0\nn\\
d=10: && [01000]+[00000]\label{repJ}
\eea
 They realize the Lorentz $J_{mn}$ and ghost currents $\Delta$ in $d=4,6,10$.
 In addition the extra singlet in $d=4$ represents the axial current $J_5$ while in $d=6$
  the triplet realizes the $Sp(1)$ ${\cal R}$-symmetry current $J_{(ij)}$.

  The contribution of $\lambda,w$ to
the partition function
 (\ref{zpart}) can then be computed by counting polynomials
 $\lambda_0^n, \lambda_1 \lambda_0^n ,w_1 \l_0^n$ with bispinor
 products restricted according to  (\ref{prod1},\ref{repJ}).
 Finally the total partition function follows
by multiplying the pure spinor result 
with the
contribution coming from the free
$\t,p,x$-system:
\bea
Z_{\theta, p,x}(q|t)&=& (1-t)^{\bf S_d}\,
\,\prod_{n=1}^\infty   {
(1-q^n \,t^{1})^{\bf S_d}\,(1-q^n \,t^{3})^{\bf \bar{S}_d}
\over
(1-q^n \,t^{2})^{\bf V_d} }
\label{zns}
 \eea
 As we mentioned before here and below we omit the
 contribution $(1-t^2)^{-{\bf V_d}}$ coming from the bosonic zero modes $x_0^m$,
 i.e. we consider spacetime constant fields.
 The total partition function will be then written as:
\be
 Z(q|t)=Z_{\l,w}(q|t)\,Z_{\theta, p,x}(q|t)=Z_0(t)+q\, Z_1(t)+\ldots
\ee
 The finite polynomials $Z_{\ell}(t)$ encode the informations about multiplicities
 and charges of string states. Aim of this work is to evaluate $Z_{0,1}(t)$
 for the massless and first massive string state in
 $d=4,6,10$.

Some comments about the relation between the string partition function and the
Q-cohomology of the corresponding string theory are in order.
First the string partition sums over all ghost number states while the cohomology
analysis here will be often restricted to ghost number one. The agreement between
the twos does not imply that higher spin cohomology is empty (although this is
mostly the case) but only that higher ghost number states, if exist, come
in field/antifield pairs.
Second the string partition function counts states off-shell at a
fixed momentum while states in the cohomology of $Q$ are often on-shell\footnote{J.F.M.
thanks R. Russo for discussions on this point}.
This implies in particular that the cohomology of $Q$ at zero momenta
is empty every time massive states come on-shell since $(\partial^2+m^2)\phi=0$
implies $\phi=0$ if $p=0$.
 This will be confirmed by the
 computation of the string partition function at the first massive level
  in $d=6,10$ where states come always in worlsheet bosonic/fermionic pairs
  and the resulting partition function cancels.
 We stress the fact that this cancelation is not related to supersymmetry since we
 are counting states keeping track of their $SO(d)$ representations,
 it is a cancelation between fields and antifields.
  In the pure spinor formalism it is hard to separate the two contributions
  without spoil $SO(d)$ covariance.
  As we will see physical states in $d=6,10$ at the first massive level
  can be isolated by keeping the contributions from the modes $\l_1,\t_1$ separated
  from those coming from their momenta $p_1,x_1,w_1$.
   The resulting spectrum organizes into a massive on-shell spin-two multiplet
   of the minimal
  supersymmetry in $d=6,10$.
 In $d=4$  the massive multiplet comes off-shell and the string and cohomology results
 therefore agree. This is also the case for massless states in $d=4,6,10$.

  In addition we will show how that the central charges of the Virasoro
algebra coming from the partition function of pure spinors
match the naive counting above in $d=4,6,10$. To this
 purpose one rewrites the massless partition function
 of pure spinors $Z_{0}$ in terms of a free system
of infinitely many fields \cite{Berkovits:2005hy}:
\be
Z_{0}(t)= \prod_{n=1}^\infty (1- t^n)^{-N_n}
\ee
 with some $N_n$. In terms of this free description
the central charge of the Virasoro current
 can be read from the logarithmic divergent term in
the small $x$ expansion of $\log Z_{0}(t=e^x)$:
 \bea
 -\log Z_{0}(e^x) &=&\log(x)\sum_n \,N_n+\ldots=\ft12 c_{\rm vir}\, \log(x)+
\ldots    \label{centr}
\eea
 In $d$-dimensions the theory is critical if the central charge of the
 $(\l,\t,p,w)$-system computed in this way cancel that of the free bosons,
 i.e. if $c_{\rm vir}=-d$.
 Notice that in order to do this computation we do not need
to determine $N_n$, but simply expand the left hand side of this
expression and find the coefficient of $\log x$ as
$\ft12\, c_{\rm vir}$. The results will be shown in agreement
with the naive counting of degrees of freedom of pure spinors in
$d=4,6,10$.

\subsection{$d=4$ }
\label{ss4}

  At the massless level the pure spinor constraint in $D=4$ can be written as:
 \bea
&& \lambda^\a_0 \, \bar{\lambda}^{\dot{\a}}_0=0\label{pure4}
\eea
 with $\l_0^A=\{ \l_0^\a, \l_0^{\dot \a}\}$ and $\a,\dot{\a}=1,2$.
 States in $d=4$ will be labelled by their representations under the
  $SO(4)\sim SU(2)\times SU(2)$ Lorentz group and the $U(1)$ charge.
  The $U(1)$ charge will be traced by the powers of 
  $t$, while Lorentz quantum
  numbers will be labelled  by the $[j_1,j_2]$ spins.
We introduce the short hand notation:
\be
 \lambda_0^\a: {\bf s}\, t  \equiv [\ft12,0]\,t \quad\quad
 \bar{\l}_0^{\dot{\a}}: {\bf c}\, t\equiv [\ft12,0]\,t
\ee
Polynomials satisfying (\ref{pure4}) are built of
 symmetric combinations of either $\lambda_0^\a$ or $\bar{\l}_0^{\dot{\a}}$:
 \be
 \l_0^n:\quad\quad  ([\ft{n}{2},0]+[0,\ft{n}{2}])\, t^n \label{l4}
 \ee
 In particular all products containing both ${\bf s}$ and ${\bf c}$
representations have
 been suppressed in (\ref{l4}).
Here and below we will often suppressed Lorentz indices when write
polynomials $\l_0^n$ of pure spinors. Complete symmetrization between
the $\l^A$ will be always understood.

 The massless partition function follows then by
 summing up (\ref{l4}) over $n$ and multiplying by the free
 $\theta_0$ -contribution $(1-t)^{{\bf s}+{\bf c}}$.
 \bea
 Z_0(t) &=& {\rm tr}_{{\cal H}_0}
 \, (-)^F\, q^{L_0} \, t^J=(1-t)^{{\bf s}+{\bf c}}\left[ {1\over (1-t)^{\bf s}}+
 {1\over (1-t)^{\bf c}}-1\right]\nn\\
 &=& 1-{\bf s}\,{\bf c}\, t^2+({\bf s}+{\bf c})\, t^3-t^4\nn\\
&=& 1-4 \, t^2+4 t^3-t^4 \label{p40}
\eea
 The two series in the bracket come from symmetric polynomials
 of $\l_0^\a={\bf s}$ and $\bar{\l}_0^{\dot{\a}}={\bf c}$
 respectively while the minus one subtracts the overcounted identity.
 The last line display the blind dimensions. It is important to notice that
the contribution of $\theta_0$ cancels exactly the denominators (\ref{p40})
leaving a finite polynomial $Z_0(t)$. This will be always the case in
any dimension.

 Even and odd powers of $t$ in (\ref{p40}) correspond to spacetime
 bosonic and fermionic degrees of freedom respectively.
 The polynomial $P_0(t)$ describes the off degrees of freedom
of a massless ${\cal N}=1$  vector multiplet in $d=4$ with content
$(A_\mu-\Lambda;\psi_\alpha,\bar{\psi}_{\dot{\a}};
 D)$ with $\Lambda$ parametrizing the gauge invariance. Notice that the
 powers of $t$ describe precisely twice the dimensions of these fields, i.e.
 $0$ for the gauge parameter, $1$ for the vector, $3/2$ for the gaugino and
 $2$ for the auxiliary field $D$.

Now let us consider the first massive string level.
 We have an extra pure spinor constraint
 and gauge invariance:
 \bea
&& \lambda^\a_0 \, \bar{\lambda}^{\dot{\a}}_1
+\lambda^\a_1 \, \bar{\lambda}^{\dot{\a}}_0=0\nn\\
&&\delta w_{1 \a} = \Lambda_{1\a\dot{\a}} \bar \l^{\dot{\a}}_0 \,,
~~~~ \delta\bar w_{1\dot{\a}} =
\Lambda_{1\a\dot{\a}} \l^\a_0\label{cont4}
\eea
 It is important to notice that when combined with (\ref{pure4}), eq. (\ref{cont4})  implies
\be\label{cic}
 \lambda^\a_0 \, \bar{\lambda}^{\dot{\a}}_1= \lambda^\a_1 \, \bar{\lambda}^{\dot{\a}}_0=0
\ee
therefore operators satisfying the pure spinor constraint
contain either $\l_{0,1}^\a$  or $\l_{0,1}^{\da}$. Invariance under (\ref{cont4})
implies that $w_{1A}$ appear only in the combination $w_{1\a}\l^\b$ and
 $w_{1\da}\l^{\db}$. In appendix B, we derive eq.~(\ref{cic}) by 
 using covariant equations. 

 The $SU(2)^2\times U(1)$ content of
 pure spinor states satisfying (\ref{cont4}) is then given by:
\bea
&& \lambda_1 \lambda_0^n: \quad\quad
\left(
 [\ft12, 0]\times [\ft{n}{2}, 0]+  [0,\ft12]\times [0,\ft{n}{2}\right)\, t^{n+1}  \nn\\
&& w_1 \lambda_0^{n} : \quad\quad\left(
 [\ft12, 0]\times [\ft{n}{2}, 0]+  [0,\ft12]\times [0,\ft{n}{2}]\right)\,
 t^{n+3}  \quad\quad n\geq 1
\label{z14s}
\eea
The total contribution coming with $\l_1,w_1$ modes follows then
by summing up over $n$ and multiplying by the contribution $(1-t)^{{\bf s}+{\bf c}}$
coming from $\theta_0$'s. One finds:
\bea
Z_{\l_1,w_1} &=& (1-t)^{{\bf s}+{\bf c}}\left( {{\bf s}(t+t^3)\over (1-t)^{\bf s}}+ {{\bf c}(t+t^3)
\over (1-t)^{\bf c}}-({\bf s}+{\bf c})t^3\right)\label{lw4}
\eea
Finally one should add the contributions coming from $\t_1,p_1,x_1$:
\be
Z_{\t_1,x_1,p_1} = \left[{\bf s}\, {\bf c}\, t^2-({\bf s}+{\bf c})\,(t+t^3)\right]Z_0(t) \label{pxt4}
\ee
with $Z_0(t)$ given by (\ref{p40}). Summing (\ref{pxt4}) and  (\ref{lw4}) one finally
finds: \be Z_1(t)={\bf s}\, {\bf c}\,(1-t)^{{\bf s}+{\bf c}}
\label{z14}\ee
 Even and odd powers of $t$ correspond to spacetime bosonic and fermionic degrees of
freedom  respectively. By expanding (\ref{z14}) one finds the bosonic field content
$(g_{\mu\nu},b_{\mu\nu},4 A_\mu,\varphi)$. This multiplet is generated by
acting will all supercharges on a vector field, i.e. it has 
$4\times 2^4$ degrees of freedom, 
and contains as a highest 
helicity state a spin two particle.

 Finally we can compute the central charge of the $d=4$ system.
 Plugging the massless partition function for pure spinors
$Z_{\l_0}$ in (\ref{centr}) one finds
 \bea
 -\log Z_{0}(t)(e^x) &=&  -2 \log(x)+\ldots
\eea
leading to $c_{\rm vir}=-4$, therefore the theory is critical in $d=4$ !

\subsection{$d=6$ }

The massless pure spinor constraint in $d=6$ can be written as:
\be
\lambda^{[A}_{0i}\,\lambda^{B]i}_{0}=0
\ee
with  $A=1,..4$, $i=1,2$, $\mu=1,..6$.
 This implies that in the symmetric product of $n$ $\lambda_0$'s
 only the representation
 \be
 \l_0^n: \quad\quad \lambda^{(A_1}_{0(i_1}\,\ldots \,\lambda^{A_n)}_{0i_n)}
 = [00n]_{n\over 2}\, t^n
\ee
 will survive. Here $[n_1 n_2 n_3]_j$ denote
 the $SO(6)$ dynkin labels and $SU(2)$ spin. In this notation
  ${\bf S_d}=[001]_{1\over 2}$.
 The massless partition function can then be written as
 \bea
 Z_0(t) &=& (1-t)^{\bf S_d}\, \sum_{n=0}^\infty [00n]_{n\over 2}\, t^n\nn\\
 &=& 1-[100]_0\, t^2+[001]_{1\over 2}\, t^3-[000]_1\, t^4\nn\\
 &=& 1-6 \, t^2+8 t^3- 3 t^4
 \eea
The cohomology $Z_0(t)$ now describes the off degrees of freedom
of a massless ${\cal N}=1$  vector multiplet in $d=6$,
 i.e. $(A_\mu-\Lambda;\psi^A_i;
 D_{(ij)})$ with $i,j=1,2$.

 At the first string level one has an extra constraint and a gauge invariance:
\bea
&& \lambda^{[A}_{0i}\,\lambda^{B]i}_{1}=0\nn\\
&& \delta w_{1A i} =\Lambda_{1[AB]} \lambda_{0 i}^B
\eea
Gauge invariant states satisfying the pure spinor constraints are
built in terms of the following monomials:
 \bea
&& \lambda_1 \lambda_0^n: \quad\quad\left(
 [00,n+1]_{n\pm 1\over 2}+[10,n-1]_{n+1\over 2}\right)\, t^{n+1}  \nn\\
 && \t_1 \lambda_0^n : \quad\quad -[001]_{1\over 2}\times
[00n]_{n\over 2}\,t^{n+1}\nn\\
&& w_1 \lambda_0^{n} : \quad\quad
\left( [0,1, n]_{n-1\over 2}+[0,0,n-1]_{n\pm 1\over 2} \right)\,t^{n+3}
 \quad\quad n>0 \nn\\
&& x_1 \lambda_0^n : \quad\quad [100]\times
[00n]_{n\over 2}\,t^{n+2}\nn\\
&& p_1 \lambda_0^n : \quad\quad -[001]_{1\over 2}\times
[00n]_{n\over 2}\,t^{n+3}\nn
 \eea
In addition one has an extra $(1-t)^{\bf 8}$ coming from powers of $\theta^{Ai}$.
Collecting the various contributions one finds
$q\, P^{\rm phys}_1(t)=Z_{\l_1,\t_1}(t)=-Z_{w_1,p_1,x_1}(t)$ with :
\bea
P^{\rm phys}_1(t) &=& -t^{2}\, [1,0,0]_{0}
+t^{3}\,[0,1,0]_{1\over 2}+ t^4\,([2,0,0]_0-[0,0,0]_{1})\nn\\
&&-t^5\,[1,1,0]_{1\over 2} +t^6\, ([0,2,0]_0+[1,0,0]_{1})
- t^7\,[0,1,0]_{1\over 2}+t^8\, [0,0,0]_0\nn\\
&=& - 6 \, {t}^{2}
 +8 \, {t}^{3}+17\, t^4
 -40 \, t^5+28 \, {t}^{6}-8 \, {t}^{7}+{t}^{8}
 \eea
 The polynomial $P^{\rm phys}_1(t)$ describes the degrees of freedom of
 a massive spin two multiplet in $d=6$:
$(g_{\mu\nu}-\Lambda_\mu;\psi_{\mu A}^i-\Lambda_A^i; C_{\mu\nu\rho}^+,C^{(ij)}_\mu
-\Lambda^{(ij)};
\lambda^i_A ;\varphi )$, see Appendix \ref{secmult} for the construction
of the massive supermultiplet.

  For the central charges one finds
 \bea
-\log Z_{0}(t)(e^x)  &=& -3 \log(x)+\ldots
\eea
leading to $c_{\rm vir}=-6$ and therefore the theory is critical in $d=6$ !

\subsection{$d=10$ }

 The pure spinor condition in $d=10$ reads:
 \be
 \lambda_0 \gamma^m \lambda_0 =0 \label{pure10}
 \ee
  This condition restrict the
 symmetric products of pure spinors $\lambda_0$ to the representation:
\be
\lambda_0^n: \quad\quad [0000n]\, t^n
\ee
 Summing up over $n$ one finds
 \bea
 Z_0(t) &=&(1-t)^{\bf 16}\, \sum_{n=0}^\infty [0000n]\, t^n\nn\\
 &=&  1-{\bf 10_v}\, t^2+{\bf 16_s}\, t^3-
{\bf 16_c}\, t^5+{\bf 10_v}\, t^6- t^8
\eea
The cohomology $Z_0(t)$ now describes
 the on degrees of freedom of a massless ${\cal N}=1$
 vector multiplet in $d=10$, i.e. $(A_\mu-\Lambda,\psi^\alpha)$
 and their antifields.

As usual at level one one has new constraints and gauge invariances:
\bea
&& \lambda_0 \gamma^\mu \lambda_1=0\nn\\
&& \delta w_{1\a} = \Lambda_{1\mu} (\g^{\mu} \l)_{0\a}
\eea
 This restricts the allowed monomials to the representations:
\bea
&& \lambda_1 \lambda_0^n: \quad\quad\left(
  [0000,n+1]+[0010,n-1]\right) \, t^{n+1} \nn\\
 && \t_1 \lambda_0^n : \quad\quad -[00001]\times
[0000n]\,t^{n+1}\nn\\
&& w_1 \lambda_0^{n} : \quad\quad
\left( [0100,n-1]+[0000,n-1] \right)\,t^{n+3}
 \quad\quad n>0 \nn\\
&& x_1 \lambda_0^n : \quad\quad [10000]\times
[0000n]\,t^{n+2}\nn\\
&& p_1 \lambda_0^n : \quad\quad -
[00010]\times
[0000n]\,t^{n+3}
 \eea
times $\theta_\alpha$'s contributing an extra $(1-t)^{\bf 16_s}$.

Collecting all the pieces and summing up over $n$ one finds
$q\, P^{\rm phys}_1(t)=Z_{\l_1,\t_1}(t)=-Z_{w_1,p_1,x_1}(t)$ with :
\bea
P^{\rm phys}_1(t) &=&  -t^2\, [10000]+t^3\, [00010]+t^4\, [20000]
-t^5\, ([00001]+[10010])\nn\\
&& +t^6\, ([00100]+[10000])-t^8\,
([00000]+[01000])+t^9\, [00001]\nn\\
&=&- 10 \, {t}^{2} +16 \, {t}^{3}+54 \, {t}^{4}
-\hbox{160} \, {t}^{5}+\hbox{130} \, {t}^{6}-46 \, {t}^{8}+16 \, {t}^{9}
 \eea
This is precisely the content of a massive spin two multiplet in $D=10$:
$$
(g_{\mu\nu}-\Lambda_\mu;\psi_{\mu\dot{\alpha}}-\Lambda_{\dot{\alpha}},
\lambda_\alpha; C_{\mu\nu\sigma}-\Lambda_{\mu\nu},A_\mu-\Lambda;
)
$$
 with $128_B-128_F$ physical degrees of freedom.

For the central charges one finds
 \bea
 -\log Z_{0}(t)(e^x)  &=& - 5\log(x)+\ldots
\eea
leading to $c_{\rm vir}=-10$ and therefore the theory is critical in $d=10$ !

\section{Cohomology}
\label{seccohomology}

In the present section we derive the general form of the
field equations coming from $Q$-invariance at ghost number one.
The solution for $d=4$ is worked out in full details.
The case $d=10$ has been already discussed in
\cite{Berkovits:2002qx}.

\subsection{General results}

Physical states are in one-to-one correspondence with states
in the cohomology of the BRST operator
\be
Q = \int dz \l^{A} d_{A}
\quad\quad
  d_A=p_A-\ft12 \partial x_m \gamma^m \theta-
\ft18 \gamma^m \theta \theta \gamma_m \partial \theta
\ee
By $\int$ we will always mean ${1\over 2 \pi \a' i} \oint$. The operator
generates a symmetry of the lagrangian and is nilpotent if the constraint
(\ref{pure}) is satisfied.
The set of rules that we are using are recalled for convenience.
The fundamental fields have been already introduced
in the previous sections. In terms of these fields,
one defined the composite
operators $\Pi_{m}=\partial x^m+\ft12 \theta \gamma^m \partial \theta$,
$d_{A}, \p \theta^{A}$ (see \cite{Grassi:2005sb,Wyllard:2005fh} for the
conventions in $d=4,6$)
which satisfy the affine Kac-Moody algebra
\bea\label{KM}
&&
d_{A}(z) d_{B}(w) \rightarrow -{\a'\over (z-w) } \,\g^{m}_{AB} \Pi_{m}(w) \,, ~~~
d_{A}(z) \Pi_m(w) \rightarrow{\a'\over (z-w)} \,\g_{m, AB} \p\theta^{B}(w) \,,
\nonumber\\
&&
\Pi_{m}(z) \Pi_{n}(w) \rightarrow - {\a'\over (z-w)^2}\, \eta_{mn}\,, ~~~~~~
d_{A}(z) \p\t^{B}(w) \rightarrow {\a'\over (z-w)^2} \,\delta_{A}^{~B}\,,
\eea
The BRST transformations of the fields read
\bea\label{brstrules}
&&
Q x^{m} = \ft12 \l^{A} \g^{m}_{AB} \theta^{B}\,, ~~~~~~~
Q \t^{A} = \l^{A}\,, ~~~~~~
Q \Pi_{m} =  \l_{A} \g^{m}_{AB} \p\theta^{B}\,, \nonumber\\
&& Q d_{A} = -\Pi_{m} \g^{m}_{AB} \l^{B}\,, ~~~~~~~ Q \l^{A} =0\,,
~~~~~~ Q w_{A} =- d_{A}\,, \eea The BRST transformation rules are
nilpontent up to gauge variations. Indeed applying twice the BRST
charge on $w_{A}$ we get $Q^{2} w_{A} = \Pi_{m} \g^{m}_{AB}
\l^{B}=\delta_{\Pi_m} w_A$ with
$\delta_\Lambda w_{A} = \Lambda_{m} \g^{m}_{AB} \l^{B}$.
Therefore, the only vertex operators which are admissible are
those which are gauge invariant under the gauge transformations
for the $w$'s.

Physical states appear in the cohomology of the BRST  operator $Q$
at ghost number $n_\l-n_w=1$. We denote string vertices  by
$U_\ell^{(q)}$ with $\ell$ labelling the string level and $q$ the
ghost charge. The vertices $U^{(q)}_\ell$ are defined up to gauge
transformations
 $\delta U^{(q)}_\ell=Q U^{(q-1)}_\ell$.
 At the massless level one finds:
 \bea
U^{(1)}_0 &=& :\l^{A} A_{A}:  \\
U^{(0)}_0 &=& : \Omega: \nn \eea with $A_{A},\Omega$
arbitrary superfields\footnote{ Normal order here and below
follows the definition \cite{DiFrancesco:1997nk} $
:A B:(z) = {1\over 2\pi i}\,\oint {dy \over (y-z)} A(y) B(z)\, $
and $:A B C:(z) = {1\over (2\pi i)^2}\,\lim_{w\rightarrow z}
\Big[\oint_{{\cal C}_{z}} + \oint_{{\cal C}_{w}} \Big] {dy \over
(y-w)} A(y) B(w) C(z)$.}.
 On a generic superfield $A(x,\theta)$ the BRST operator $Q$ acts as
 a supersymmetric derivative :
 \be
 Q A=\l^A \,D_A A
 \ee
Acting with $Q$ and imposing $Q
U^{(1)}_0=0$, $\delta U^{(1)}_0=Q U^{(0)}_0$ one finds 
\be
D_{(A}A_{B)}= \g^{m}_{AB} A_{m}\,,
\quad\quad \delta A_{A}=D_A \Omega\,, ~~~~~
\delta A_{m} = \p_{m} \Omega\,,
 \ee It is not hard
to see that this
gives the degrees of freedom of a vector multiplet in $d$
dimensions.

 Let us consider now the first massive string level.
For the ghost number 0,1 vertices one finds: \bea\label{verA}
U^{(0)}_{1} &=& :\p \t^{A} \Omega_{A}: + :\Pi^{m} \Gamma_{m}: +
:d_{A} \Lambda^{A}: + :J_{A}{}^B \Phi_B^A:\,\\
U^{(1)}_{1} &=& :\p\l^{A} A_{A}: + :\l^{A} \p \t^{B} B_{AB}: +
:\l^{A} d_{B} C^{B}_{A}: + :\l^{A} \Pi^{m} H_{A m}:  +
:J_{A}{}^B \l^{C}\,  F^{A}_{CB}:\nn
 \eea Again
superfields $A_{A}, B_{AB},...\Omega,\Gamma,..$ are functions of
the (super)spacetime coordinates $(x^m,\theta^A)$. We denote by
$J_{A}{}^B$ the gauge invariant combinations \be\label{combA}
J_A{}^B =
{\cal P}_{AC}^{BD} :w_D \l^C:\,,
\ee
 with ${\cal P}_{AC}^{BD}$ a projector into the gauge invariant components
 satisfying $\delta J_A{}^B=0$.
For example in  $d=4$, $J_A{}^B= J\delta_A^B+J_{mn}
\gamma^{mnB}_A+ J_{5}\gamma_{5A}^B$, where $J$ is the ghost
current, $J_{mn}$ are the Lorentz generators in  the pure spinor
sector and $J_{5}$ generates chiral rotations of the pure spinors.
In a similar way the Lorentz content of $J_A{}^B$ in $d=6,10$ is
given in (\ref{repJ}).

In taking products of $J_A{}^{B}$ and $\l^A$ one should take
particular care with normal orderings. In particular, 
two operators that are independent at the classical level
can mix under normal
ordering. The first few of these relations take the form:
\bea\label{anA} \Xi^{A}_{BC} J_{A}^B \l^{C}=0 &\Rightarrow&
\Xi^{A}_{BC} :J_{A}^B \l^{C}:=
-\alpha'    \Xi^{A}_{BA} \p \l^{B} \,,   \\
K^{A}_{BCD} J_{A}^B \l^{C} \l^D=0&\Rightarrow& K^{A}_{BCD}
:J_{A}{}^B \l^{C} \l^D:= -2\alpha'  K^{A}_{B(CA)}\lambda^{C}\p
\l^{B} \nn \eea
The right hand side here refers to product of functions rather than operators.
This implies that a relation  that hold at the classical level
$\alpha'\to 0$ due to pure spinor constraints can fail in
the full quantum theory.
 We will refer to these relations as anomalous
Ward identities. In particular the first of these equations implies
that operators $:\p \l^A:$ and  $:J_{A}{}^B \l^{C}:$ appearing in
$U^{(1)}_1$ are not independent, i.e.
 the vertex operator $U^{(1)}_{1}$ is defined up to
the algebraic gauge transformations \be\label{verC} \delta
F^{A}_{BC} = \Xi^{A}_{BC}\,, ~~~~~~~ \delta A_{A} =- \alpha'
\Xi^{B}_{AB} \ee where $\Xi_{BC}^A$ is a gauge parameter
satisfying (\ref{anA}).
  These gauge transformations can be used to restrict $F^A_{BC}$ to those
components satisfying $F^A_{BC} J_A^B \lambda^C\neq 0$ at the
classical level. Here will always work in this gauge.
 In a similar way
$K^A_{BCD}$ reflects a relation between operators appearing at
ghost number 2 and will be important in our analysis below. The
Lorentz content of $\Xi^{A}_{BC}$ and  $K^{A}_{BCD}$ depends on
the dimensions and will be worked out below in some relevant
cases.

Acting with the BRST charge on the vertex operator $U^{(1)}_{1}$,
one obtains the following equations of motion \bea &&\label{vA}
\l^{A} \l^{B} \p\t^{C} \Big( D_{(A} B_{B)C} -  \g^{s}_{C(A}
H_{B)s} \Big) =0\,,
\\&&\nonumber\\
&&\label{vB} \l^{A} \l^{B} \Pi_{s} \Big( D_{(A} H_{B)}^{s} -
\g^{s}_{C(A} C_{B)}^{C} \Big)=0\,,
\\&&\nonumber\\
&&\label{vC} \l^{A} \l^{B} d_{C} \Big( D_{(A} C_{B)}^{C} +
F^{C}_{(AB)} \Big)= 0 \,,
\\&&\nonumber\\
&& \label{vD} :\l^{C} \l^{D} J_{A}{}^B: \Big( D_{D} F_{C B}^A -
K^A_{BCD} \Big) =0\,,
\\&&\nonumber\\
&&\label{vE} \l^{A} \p\l^{B} \Big( D_{A} A_{B} + B_{AB} + \alpha'
\g^{m}_{BC } \p_{m} C_{A}^{C} -2 \alpha' D_{(C} F^{C}_{A) B} +
2\alpha'  K^C_{B(AC)} \Big)= 0 \eea
   The contributions
of $K^A_{BCD}$ to (\ref{vD}) and (\ref{vE}) cancel against each
other in $Q U^{(1)}_1$ according to (\ref{anA}) and allow us to
treat these two equations independently.
 It is important to notice that the introduction of this superfield
ensures the gauge invariance under the symmetry (\ref{verC}) of
the equations of motion~(\ref{vA})-(\ref{vE}). Taking for example
eq. (\ref{vD}) and by performing the gauge transformation
(\ref{verC}), one can see that the compensating gauge
transformation is \be\label{comA} \delta K^A_{BCD} = D_{D}
\Xi_{CB}^A \ee In addition, the equations of motion are invariant
under the gauge transformations \bea\label{verD} \delta A_{A} &=&
\Omega_{A} + \alpha' \g^{m}_{AB} \p_{m} \Lambda^{B} - \alpha'
D_{B} \Phi_A^{B}\,,
\nonumber \\
\delta B_{AB} & = & - D_{A} \Omega_{B} +  \g^{m}_{AB} \Gamma_{m}
\,,
\nonumber \\
\delta H_{A m} &= & D_{A} \G_{m} - \g_{m, AB} \Lambda^{B}
\nonumber \\
\delta C^{A}_{B} &= & - D_{A} \Lambda^{B} -\Phi^A_B \,,
\nonumber \\
\delta F^{A}_{CB} &= & D_{C} \Phi^A_B  \,, \eea These gauge
transformations are needed in order to select the physical states
and they are used to set some of the field to zero. Indeed the
gauge parameters $\Omega_A,\Gamma_m,\Lambda^A,\Phi_A^B$ can be
used to set \be A_A=\gamma^{mAB}B_{AB}=\gamma^{mAB} H_{mB}={\cal
P}_{AC}^{BD}\,C_B^A=0 \label{gaugefix} \ee respectively. The
resulting fields should be plugged into the BRST equations
(\ref{vA}-\ref{vE}). Formally the first four equations
 can be written as: $H=dB$, $C=dH$, $F=dC$,
$K=dF$, projected on specific representations carried by the
worldsheet operators multiplying them. These equations, as we will
see,  allow us to express all fields in terms of a single
superfield $B_{[mnp]}$.
Below we will show this for $d=4$. The case $d=10$ has been worked
out in \cite{Berkovits:2002qx}. There is an important difference
between $d=4$ and $d=6,10$. In $d=4$ we will find that the massive
fields are off-shell while in $d=6,10$ are on-shell. This is due
to the fact that in $d=4$ there is no enough supersymetries to
build the Laplacian operator out of rasing supercharges Q's.


\subsection{$d=4$  }

We use the following notation for Dirac matrices: $ \g^{mnp..}$
for antisymmetrized combinations of the Dirac matrices $\g^{m}$. A
symmetric bispinor $\psi^{(AB)}$ can be decomposed as follows
$\psi^{(AB)} = \psi^{m} \g^{(AB)}_{m} + \psi^{mn} \g^{(AB)}_{mn}$;
an antisymmetric bispinors $\psi^{[AB]} = \psi C^{[AB]} +
\psi^{mnp} \g_{mnp}^{ [AB]} + \psi^{mnpq} \g^{[AB]}_{mnpq}$. The
indices are raised and lowered with the antisymmetric tensor
$C^{[AB]}$.

The pure spinors are represented by a Dirac spinor $\l^{A}$
($A=1,\dots, 4$) and the pure spinor constraint is $\l \g^{m} \l
=0$. As a warming-up exercise, we compute the cohomology at
massless level. The most general massless vertex operator is
\bea\label{fourD} U^{(1)}_{0} = \l^{A} A_{A}(x,\t) \eea and the
gauge symmetry is given by the scalar superfield $\Omega$. The
BRST condition implies $\l^{A} \l^{B} D_{A} A_{B} =0$ up to gauge
transformations $\delta A_{A} = D_{A} \Omega$. Therefore, the most
general solution is given by \be\label{fourE} U^{(1)}_{0} =
\l^{A} \left( D_{A} M + \g^{5}_{AB} D^{B} M_{5} \right) \,. \ee
The first term can be removed by a gauge transformation, but the
second term represents an element of the cohomology. Notice that
we have to require the reality condition in order not to spoil
this symmetry of the theory. This implies that the dofs are
represented by a real scalar superfield $M_{5}$. This computation
appeared also in \cite{Grassi:2005sb}.

The BRST condition on the physical states gives the equations of
motion in (\ref{vA})-(\ref{vE}). However, one has to remove the
factors in front of the equations and to project the equation
along the pure spinor directions. In the $d=4$, a symmetric
bispinor is decomposed as $\l^{A}\l^B = {1\over 4} \g^{AB}_{m}
\psi^{m} + {1\over 6} \g^{AB}_{mn} \psi^{mn}$ and $\psi^{m} =0$
because of the pure spinor condition. This implies for example
that the first equation (\ref{vA}) becomes \bea\label{fourF}
\g_{mn}^{AB} \left( D_{(A} B_{B)C} -  \g^{s}_{C(A} H_{B)s}
\right) =0\, , \eea and so on. It is undoubtedly convenient to use
$SU(2)\times SU(2)$ (Weyl)
 indices to label $SO(4)$ representations.
In terms of these definitions the superfields decompose as follows
\bea\label{pllA} &&
A_{A} =\{A_{\a}, A_{\dot \a}\}\,, ~~~~ \nonumber \\
&&
B_{AB} = \{B_{\a\b}, B_{\a\db}, B_{\da\b}, B_{\da\db}\}\,, ~~~~ \nonumber \\
&&
H_{A m } =\{ H_{\a \b\db}, H_{\da \db \b}\}\,, ~~~ \nonumber \\
&&
C_{A}^{~B} = \{C_{\a}^{~\b}, C_{\a}^{~\db}, C_{\da}^{~\b}, C_{\da}^{~\db}\}\,\nn\\
&& F^A_{BC} =\{ F^{\a}_{(\b\g)}, \bar F^{\da}_{(\db\dg)}
\}\,,
\eea
 where we have used the $\Xi_{\a}^{\b \db},\Xi_{\da}^{\db \b}$-gauge symmetry
 in order to put $F^A_{BC}$ in this form.
The currents $J_A{}^B$ become $\{J_{\a\b},J_{\da,\db}\}$ and
satisfy the anomalous Ward identities (\ref{anA}) which expressed 
in Weyl indices become
 \bea\label{xfourC} && : J_{\a\b} \l^{\b}: = - \a'
\p\l_{\a}\,,
\nonumber \\
&&
: J_{\a\b} \bar\l^{\db}: = 0\,,
\nonumber \\
&&
: J_{\a\b} \l^{\b} \l^{\g}: = - \a' \p\l_{\a} \l^{\g} - \a' \delta_{\a}^{\g} \p\l_{\b} \l^{\b}\,,
\nonumber \\
&&
: J_{\a\b} \l^{\b} \bar\l^{\dg}: = 0\,, ~~~~~~~
\nonumber \\
&&
: J_{\a\b} \l^{\b} \l^{\g}: = 0\,.
\eea
and the hermitian conjugates.
 For
simplicity in what follows we omit complex conjugate equations
that can be easily found by exchanging dotted and undotted indices.
The Ward identities (\ref{xfourC}) follow from (\ref{anA}) by introducing
$K_{\b \s}$  and its complex conjugate.
 Eqs (\ref{vA}-\ref{vE}) become
 \bea \label{llpt} &&
 D_{(\a}  B_{\b)\g} =0\,, \nonumber \\
&&  D_{(\a} B_{\b)\dg}  -  H_{(\a\b) \db} = 0\,,
\eea
\bea \label{llPi} &&
 D_{(\a}  H_{\b)}{}^{\a}_{\dot \d}  -  C_{\b\dot{\d}} =0\,,
\eea
\bea \label{lld} &&
 D_{(\a}  C_{\b)}^{~\g} +  F^{\g}_{(\a \b)} = 0\,, \nonumber \\
&& D_{(\a} C_{\b)}^{~\dg}  = 0\,,
\eea
\bea \label{llJA} && D_{(\b}  F_{\a)\g}^{\s} -K_{\g (\a}\delta_{\b)}^{\s}=0
\,,
\eea
\bea \label{lplA} && D_{\a} A_{\b} + B_{\a\b}
+ \a' \p_{\b\db} C_{\a}^{~\db}-2 \a' D_{(\tau} F_{\a)\b}^{\tau} +
 \a' K_{\a \b }=0 \,,
\eea

These equations are gauge invariant with respect to the gauge
symmetry (\ref{verD}) that in Weyl indices become
\be\label{xgautraA}
\begin{array}{ll}
\d A_{\a} = \Omega_{\a} - \a' D_{\g}\phi^{\g}_{~\a} +
\a' \p_{\a\db} \bar \Lambda^{\db}\,, 
\\
\d B_{\a\b} = - D_{\a}\Omega_{\b}\,, 
\\
\d B_{\db \a} = - \bar D_{\db} \Omega_{\a} + \Gamma_{\a \db}\,,
\\
\d H_{\a \b\db} = D_{\a} \Gamma_{\b\db} - \e_{\a\b} 
\bar \Lambda_{\db}\,,
\\
\d C_{\a}^{~\b} =- \phi_{\a}^{~\b} - D_{\a} \Lambda^{\b}\,, 
\\
\d C_{\da}^{~\b} = - \bar D_{\da} \Lambda^{\b}\,, 
\\
\d F^{(\a\g)}_{\b} = D_{\b} \phi^{(\a\g)}\,.  
\end{array}
\ee

The gauge symmetries
$\Omega_\a,\Gamma_{\a\db},\Lambda^{\g},\Phi_{\a}^{\b}$
can be used to set \be
A_\a=B_{\a\db}+B_{\db \a}=H^\a{}_{\a \db}=C_{\a}^{\b}=
F_{\a\g}^{\a} =0 \ee and similar for their
complex conjugates.


Now it is easy to solve the equations. From
(\ref{llpt}-\ref{llJA}), one finds \bea\label{solB}
&& B_{\a\b} = D_{\a} T_{\b}\,, \nonumber \\
&& H_{\a\b \dg} =  D_{(\a} B_{\b)\db}\,,\nonumber \\
\label{solF} &&C_{\b\dot \d} =
 D_{(\a}  H_{\b)}{}^{\a}_{\dot \d} \,, \nonumber \\
&&  F^{\g}_{\a \b}= K_{\a\b}= 0  \eea
Then eq. (\ref{lplA}) gives 
\be 
T_{\a} = -{2 \alpha'\over 3} \p_{\a\dot \a} D_{\s} B^{\s \dot \a}\,, ~~~~~
\ee
  Therefore the full cohomology at ghost number one can be written
  in terms of an unconstrained superfield
$B_{\a \db}-B_{\db \a}$. This is in complete agreement with the
result of the string partition function (\ref{z14}).
 Notice that in obtaining this result it is crucial that in $d=4$ $\l
\partial\bar{\l}=\bar{\l}\partial\l=0$ or in covariant form $\l
\gamma^{mnp} \partial \l=0$. This is not the case in $d=6,10$ and
therefore the last equation (\ref{vE}) gives a non-trivial Laplacian eq. for
$B_{mnp}$ in these dimensions. In all cases the multiplet starting with $B_{mnp}$
 contains a spin two particle
but in $d=6,10$ fields are on-shell and the multiplets
are shorter. This was explicitly shown in
\cite{Berkovits:2002qx} for the case of $d=10$.


\section{ Low dimensional models}
\label{seclow}

 Here we consider  pure
spinor constructions in $d=2,3$ dimensions.

\subsection{ ${\cal N}=(2,0)$ model}
\label{sec20}

 We consider the {\it pure} spinor system
 \bea\label{xddueF}
  && \l \bar \l = 0 \,, \nn\\
  && \delta w = \Lambda \bar \l \,, ~~~~ \delta\bar w =
\Lambda \l \,, \eea
with $\Lambda$ is the gauge parameter.
 As before we add the
anticommuting variables $(\t, \bar \t)$ and their conjugates $(p,
\bar p)$.
 The pure spinor constraint allows us
to set either $\l=0$ or $\bar \l =0$ and therefore the
$\l,w,\t,p$ system has naive central charge $c_{\l,w,\t,p}=2-4=-2$.
Therefore the system is critical in
$d=2$. We denote by $X,\bar{X}$ the bosonic degrees of freedom.

The covariant derivatives and supersymmetric line elements are
defined by \bea\label{xddueC}
 d = p + \bar{\t} \p x\,, ~~~ \bar d = \bar{p}\,, ~~~ \Pi
= \p x \,, ~~~
\bar \Pi = \p \bar x + \bar \t \p \t \,, ~~ 
\eea
 The two supercharges realize an ${\cal N}=(2,0)$ supersymmetry.
 We should stress that even if these definitions look asymmetrical, by
similarity transformations they can be put in a
symmetric form.
The BRST charge can be written as
\bea\label{xddueE}
  Q = \int (\l d + \bar \l \bar d)\,,
\eea
and acts as follows
\bea
\begin{array}{llll}
  Q \,\t^\a=\l^\a &\quad Q \,\bar{x}=\l \bar{\t}  &\quad Q\, d=\bar{\l}\, \Pi &\quad Q \,\bar{\Pi}=\bar{\l}\,\partial \t \\
  Q \,\l^\a=0 & \quad Q \,w_\a=-d_\a & \quad Q\, \bar{d} =-\l\, \Pi & \quad Q \,\Pi=0 \label{xbrstrules11}\\
\end{array}
\eea
 Acting on a generic superfield $A(x^m,\theta)$ one finds
\be
Q A=\l D A+\bar{\l} \bar{D} A \quad\quad
D = {\partial \over \partial \t}+\bar{\t} {\partial \over \partial \bar{x}}\quad\quad
\bar{D} = {\partial \over \partial \bar \t}
\ee
 The worldsheet theory in now invariant  under (apart from $\Delta$) the
 extra $U(1)$ symmetry defined by the charge assigments:
 \be
 J'=n_{d_\a}+n_{w_\a}+n_\Pi
\ee
 This new symmetry will be traced by the parameter $t'$.
 It is easy to compute the partition function of this model.
The pure spinor constraint requires that only polynomials
of either $(\l,w)$ or $(\bar{\l},\bar{w})$ are allowed.
One finds:
\be
Z_{\l,w}(q,t)=\left({2\over (1-t)}-1\right)
+q\,\left( {2 (t+t'  t^3)\over (1-t)}-2\,t'\, t^3\right)
+\ldots
\ee
Multiplying by the free contributions of $\t_{0,1},x_1,d_1$:
\be
Z_{\t_{0,1},x_1,p_1}=(1-t)^2 \left[ 1+q( t^2+t'\,t^2 -2  t-2 t' t^3)+\ldots \right]
\ee
one finds
 \bea
  Z_0(t) &=&  1-t^2 \nn\\
Z_1(t) &=&t' t^2\,(1-t)^2-  t^2 \, (1-t)^2\label{z11}
\eea
Interestingly enough, one finds a pair of multiplets at the first massive level
with opposite statistics and different $J'$ charge. Below we will confirm this
result by explicit analysis of the cohomology.

For the central charges one
finds
 \bea
-\log Z_{0}(t)(e^x)  &=& -\log(x)+\ldots \eea leading to $c_{\rm
vir}=-2$ therefore the theory is critical in $d=2$ !

 Now let us consider the ghost number one cohomology.
We start with the massless level.
The vertex operator is $U^{(1)}_{0} = \l A+\bar{\l} \bar{A}$ and
the gauge symmetry is generated by $U^{(0)}_{0} = \Omega$. The
BRST symmetry implies that \bea\label{xunoAA} D A=\bar{D} \bar{A}=0
\eea and its most general
solution is $ A_{\a} = D_{\a} M $
where $M$ is an arbitrary superfield.
$M$ can be gauged away using $\Omega$ and therefore there is
no cohomology at ghost number one.  Notice that at zero momentum this
is not true any longer and one finds that there is an element of
the cohomology given by the monomial $\bar{\l} \t$.
 The cohomology at zero momentum contains then the identity operator
 and a fermionic state at order $t^2$ in agreement with (\ref{z11}).

Let consider the first massive level. At ghost number zero one has the vertex:
\be
U^{(0)}_{1} = :\p \t^{\a} \,\Omega_{\a}: + :\Pi^{m}\, \Gamma_{m}: +
:d_{\a} \,\Lambda^{\a}: + :J_{\a}{}^\a \,\Phi_\a :\, \label{00}
\ee
 $Q U_1^{(0)}=0$ implies
 \be
J_\a^\a=\Omega_\a=\bar{\Gamma}=0\nn\\
D \Gamma+\bar{\Lambda} = \bar{D} \Gamma-\bar{\Lambda}=0
 \ee
 therefore the cohomology at ghost number zero is given in terms of a
 single unconstrained superfield $\Gamma$. The content of $\Gamma$ reproduces
 $t'$ term in (\ref{z11}).  This is an important difference with
 the case $d=4$ where the cohomology at ghost
 number zero was shown to be empty.

 The remaining states appear at ghost number one. Acting with $Q$ on (\ref{00})
 one finds the gauge transformations:
\be
\delta U^{(1)}_{1} = :\p \l^{\a}\, \Omega_{\a}: + : \bar{\l} \p \theta\, \bar{\Gamma}:
+ :\bar{\l} \Pi\, \Lambda:- : \l \Pi \,\bar{\Lambda}: - :d_{\a} \l^\a \,\Phi_\a :\,+\ldots
\label{gauge11}
\ee
with dots denoting supersymmetric derivatives of the gauge superfields.
The gauge transformations (\ref{gauge11}) can be used to set all
 components of $U^{(1)}_1$ appearing in
(\ref{gauge11}) to zero. One finds the gauge fixed vertex operator
\be
U^{(1)}_{1} =  :\l \p \bar{\t} \,\tilde{B}: + :\l \bar{\Pi} \,H:
 + :\bar{\l} \bar{\Pi} \,\bar{H}: +\left(:\l \p \t\, B:+
:\l \bar{d}\, C:  + :J \l \,  F:+{\rm h.c.} \right)
\ee
 Imposing $Q U^{(1)}_1=0$ a simple algebra leads to the following eqs.:
 \bea
 && B=\bar{B}=C=\bar{C}=F=\bar{F}=\bar{H}=0\nn\\
&& D \,\tilde{B}=D\, H=0  \label{sol11}
 \eea
 At zero momenta the degrees of freedom coming from (\ref{sol11}) are
 $\tilde{B}:-t^2(1-t)$
 and $H=t^3(1-t)$. Altogether they reproduce the $t'^0$ states in (\ref{z11}).
 It is important to stress that although the cohomology at ghost number 0,1
 already matches
 the string partition function this does not imply that the higher ghost number
 cohomology is empty. In particular there is an infinite tower of states
 $\bar{\Pi} \l^n$ that clearly belong to the cohomology,  but for $n>1$ they
  come always in pairs field/antifield with opposite statistics.

\subsection{d=3}

Next we consider a pure spinor $\l^\a$ in $d=3$ satisfying the
constraints: \be \l^{(\alpha} \l^{\beta)}=0 \label{pure3} \ee
transforming in the vector representation of $SO(3)$, i.e. the
{\bf 3} of $SU(2)$. The gauge
invariance reads \be \delta w_\a = \Lambda_{\a \b} \l^\b
\label{gauge3} \ee In addition one adds a $(p_\a,\t^\a,x^m)$
system. It is easy to list the set of
 invariant monomials satisfying (\ref{cont4}):
\bea
&&  1,\quad \l_0^\a  \nn\\
 &&  \l^\a_1,\quad  \t_1^\a,\quad \l_1^{[\a} \l_0^{\b]},\quad \t_1^\a \l_0^\b ,\quad
  x_1^m,\quad x_1^m\, \l_0^\a, \quad p_{1\a},\quad
 p_{1\a}\, \l_0^\b, \quad w_{1\a}\,\l_0^\b ,
\label{fock3} \eea
 Collecting all contributions and multiplying by $(1-t)^2$ one finds:
\bea
Z_0(t)&=& 1-{\bf 3}\, t^2+{\bf 2}\, t^3\nn\\
Z_1 (t) &=& {\bf 4}\, t^3-({\bf 3}+{\bf 5})\, t^4+ {\bf 4}\,
t^5\label{z3} \eea
Like in the $d=4$ case one finds a massless gauge multiplet in $d=3$ and
a massive long multiplet containing a spin two particle.
For the central charges one finds
 \bea
-\log Z_{0}(t)(e^x)  &=& -2\log(x)+\ldots \eea leading to
$\ft12\,c_{\rm vir}=-4$ therefore the theory is non-critical in $d=3$!

In this case it is rather easy to compute the complete spectrum at
massless and first massive level. The crucial point here in
comparison with higher dimensional examples is that the pure
spinor constraint are very strong leading to a non-trivial
cohomology only at ghost number one. Indeed we will identify all
states in the partition function result (\ref{z3}) as ghost number
1 states.

We write the pure spinor constraint as $\l \g^{m} \l =0$ and the
gauge symmetry as $\delta w_{\a} = \Lambda_{m} (\g^{m} \l)$. The
combinations $J_{\a}^{\b} = :w_{\a} \l^{\b}:$ are gauge invariant.
A bispinor $\psi^{\a\b}$ is decomposed into $\psi^{\a\b} =
\e^{\a\b} \psi + \g^{\a\b}_{m} \psi^{m}$. The Dirac matrices
$\g^{m}_{\a\b}$ are symmetric and real, they satisfies the Fiersz
identities $\g_{m, \a\b} \g^{m}_{\g\d} =0$. The notation, the action and 
the BRST symmetry are described in \cite{Grassi:2004tv}. 

The most general vertex operator of ghost number 1 at the massless
level is \be\label{treA} U^{(1)}_{0} = \l^{\a} A_{\a}(x,\t) \ee
and the gauge symmetries are represented by a scalar superfield
$\Omega$. The gauge transformations are given by $\delta A_{\a} =
D_{\a} \Omega$. Imposing the BRST invariance and using the pure
spinor condition, we see that there is no constraint on the
superfield $A_{\a}$. However, by using the gauge symmetry, we can
easily see that \bea\label{treAA} A_{\a} = a_{(\a\b)}(x) \t^{\b} +
u_{\a}(x) \t^{2}\,, ~~~~~~ \delta a_{\a\b}(x) =\p_{\a\b}
\omega(x)\,. \eea These are the dofs for an off-shell super-Yang-Mills in 3d
in agreement with (\ref{z3}).

For the massive spectrum, we consider the first massive level, by
expanding the vertex operator into conformal spin 1 worldsheet
operators.

We have that the most general vertex operator is  given by
\be\label{treB} U^{(1)}_{1} = \p\l^{\a} A_{\a} + \l^{\a} \p
\t^{\b} B_{\a\b} + \l^{\a} \Pi^{\b\g} H_{\a (\b\g)} + \l^{\a}
d_{\b} C^{\b}_{\a} + : J_{\b}^{\a} \l^{\g}: F_{\a\g}^{\g} \ee
which is invariant under the gauge symmetry \be\label{treC} \delta
F_{\a\b}^{\g} = \Xi^{\g}_{\a\b}\,, ~~~~ \delta A_{\a} = - \alpha'
\Xi_{\a \b}^{\b}\,. \ee Notice that this gauge symmetry removes
completely the field $F$, but this has an effect on the supefield
$A_{\a}$. The most general gauge transformation is generated by
the vertex operator \bea\label{treD} U^{(0)}_{1} = \p \t^{\a}
\Omega_{\a} + \Pi^{(\a\b)} \Gamma_{(\a\b)} + d_{\a} \Lambda^{\a} +
J_{\a}^{\b} \phi^{\a}_{\b} \,. \eea and the gauge transformations
are given by \bea\label{treF} \delta A_{\a} &=& \Omega_{\a} -
\alpha' \p_{\a\b} \Lambda^{\b} - \alpha' D_{\b} \phi_{\a}^{\b}
 \nonumber \\
\delta B_{\a\b} & = &
- D_{\a} \Omega_{\b} +  \Gamma_{(\a\b)} \,, \nonumber \\
\delta H_{\a (\b\g)} &= &
D_{\a} \G_{\b\g} + \e_{\a(\b}\Lambda_{\g)} \nonumber \\
\delta C^{\a}_{\b} &= & - D_{\b} \Lambda^{\a} - \phi_{\a}^{\b} \,,
\nonumber \\
\delta F_{\a\b}^{\g} &= & D_{(\a} \phi _{\b)}^{\g} \,, \eea The
gauge symmetry of $F$ is a subset of the gauge symmetry
(\ref{treC}).

By imposing the BRST symmetry, we have the only condition (the
other conditions are trivially satisfied) \be\label{treG}
\e^{\a\b} \left( D_{\b} A_{\a} + B_{\a\b} + \alpha' \p_{\g
(\a}C_{\b)}^{\g} - \alpha' D_{(\a} F_{\b)\g}^{\g} \right)=0\,, \ee
Now, we can use the gauge symmetry to removing several fields. 1)
Use $\Xi^{\a}_{\b\g}$ to remove $F^{\a}_{\b\g}$, 2) use
$\phi_{\a}^{\b}$ to remove $C_{\a}^{\b}$, 3) use $\Lambda_{\a}$ to
set $\e^{\a\b} H_{\a (\b\g)} =0$, 4) use $\Gamma_{\a\b}$ to kill
the symmetric part of $B_{\a\b}$, 5) use $\Omega_{\a}$ to set
$A_{\a}=0$. From eq. (\ref{treG}) we are left with $\e^{\a\b}
B_{\a\b} =0$. THis implies that the only physical content of the
theory stays in the superfield $H_{\a (\b\g)}$. So, the physical
states are given by the vertex operator \be\label{treH}
U^{(1)}_{1} = \l^{\a} \Pi^{\b\g} H_{\a(\b\g)}(x,\t) \ee for any
$H_{\a (\b\g)}$.  By expanding it into components, we have
\be\label{treI} H_{\a(\b\g)} = h_{\a(\b\g)}(x) + h_{\a(\b\g)
\d}(x) \, \t^{\d} + \hat h_{\a(\b\g)}(x) \, \t^{2} \ee which
contains 8 bosons and 8 fermions. Comparing the $SO(3)$ content in
(\ref{treI}) with that found in  (\ref{z3}) we find a perfect
agreement. The fields are not on-shell as to be expected by
general considerations.


\section{Conclusions and Summary}
\label{secconclusions}

In this paper we consider pure-spinor critical strings  
in dimensions $d=4,6,10$. We determine the spectrum of massless and first massive
string states in any dimension. The  results are written in terms of $SO(d)$
covariant string partition functions tracing Lorentz representations and an extra
$U(1)$ charge $\Delta$ . The Fock space is defined by string modes satisfying the pure
spinor constraint and the induced gauge invariance. This results into restrictions
on the allowed representations entering in the product of pure spinors $\l_n^A$
and/or their conjugated momentum $w_{nA}$. The outcome of this analysis
is displayed in tables 2 and 3 for the massless
and first massive level respectively.

\begin{table}[h]
\begin{center}
\begin{tabular}{|l|l|l|}
\hline
     d & Group & $\l_0^n$  \\ \hline
      4 & $SU(2)\times SU(2)$ &  $[\ft{n}{2},0]+ [0,\ft{n}{2}]$   \\
   6 & $SO(6)\times SU(2)$ &  $[0,0,n]_{\ft{n}{2}}$  \\
  10 & $SO(10)$ &  $[0,0,0,0,n]$  \\
 \hline
     \end{tabular}
\label{lambdan}
\caption{Fock space of pure spinors: The massless level.}
 \end{center}
 \end{table}

\begin{table}[h]
\begin{center}
\begin{tabular}{|l|l|l|l|}
\hline
     d & Group &  $\l_1 \,\l_0^n$  & $w_1\, \l_0^n$\\ \hline
      4 & $SU(2)\times SU(2)$   &
       $[\ft{n\pm 1}{2},0]+ [0,\ft{n\pm 1}{2}]+\delta_{n,1} [\ft12,\ft12]$&
        $[\ft{n\pm 1}{2},0]+ [0,\ft{n\pm 1}{2}]$ \\
   6 & $SO(6)\times SU(2)$   &   $[00,n\pls 1]_{\ft{n\pm 1}{2}}
   +[10,n\mis 1]_{\ft{n+ 1}{2}} $&
 $[01n]_{\ft{n- 1}{2}}+[00,n\mis 1]_{\ft{n\pm 1}{2}} $\\
  10 & $SO(10)$   &    $[0000,n\pls 1]+[0010,n\mis 1]$ & $[0100,n\mis 1]
  +[0000,n\mis 1]$\\
 \hline
     \end{tabular}
\label{lambdan1}
\caption{Fock space of pure spinors: The first massive level.}
 \end{center}
 \end{table}
 In particular specifying to $n=2$ in
table \ref{lambdan} we see that the vector components $[\ft12 \ft12]$, $[100]$
and $[10000]$ are missing in the symmetric products of two pure spinors $\l_0^A$
as expected.

The massless partition function follows by collecting the
representations in table (\ref{lambdan})
weighted by $t^{n}$, summing up over $n$ and multiplying them by the
contribution of the free fields $\theta,x,p$.  One finds
\be
 Z^d_0(t)=(1-t)^{\bf S_d}\sum_n \l_0^n\, t^n
\ee
with $SO(d)$ content
\bea
Z^{d=4}_0(t) &=& 1-{\bf 4_v}\, t^2+({\bf 2_s}+{\bf 2_c})\, t^3-t^4\nn\\
Z^{d=6}_0(t) &=&  1-{\bf 6_v}\, t^2+ 2 {\bf 4_s}\, t^3
-3\, t^4\nn\\
Z^{d=10}_0(t) &=&  1-{\bf 10_v}\, t^2+{\bf 16_s}\, t^3-
{\bf 16_c}\, t^5+{\bf 10_v}\, t^6- t^8 \label{p0s}
\eea
 In each dimension this is precisely the content of a gauge vector
supermultiplet  of minimal supersymmetry.

At the first massive states in $d=4$ one finds
\be
P^{d=4}_1(t) = -{\bf 4_v}\, t^2 (1- t)^{{\bf 2_s}+{\bf 2_c}} \label{p1s} \\
\ee
Even and odd powers of $t$ corresponnds to bosonic and fermionic degrees
of freedom respectively.
Expanding (\ref{p1s}) one fidns the degrees of freedom of a long multiplet
of minimal supersymmetry with a spin two particle as the highest 
helicity state. 
 Notice that this represents the content of a long multiplet with a
 spin two highest helicity state.
In $d=6,10$ fields come always in pairs with anti-fields and the total partition
function vanishes.
 Physical states can be identified with those coming from $\l_1,\t_1$ string modes.
 One finds $P_1(t)\equiv Z_{\l_1,\t_1}(t)=-Z_{w_1,p_1,x_1}(t)$ with
\bea
P^{d=6}_1(t) &=&  -{\bf 6_v}\, t^2+2\, {\bf 4_s}\, t^3+({\bf 20}-
 3)\,t^4-2 {\bf 20_s} t^5+({\bf 10}+3\, {\bf 6}) t^6-2 {\bf 4}_c t^7
+t^8\label{p16s}\\
P^{d=10}_1(t) &=&  -{\bf 10_v}\, t^2+{\bf 16_s}\, t^3+{\bf 54} t^4-
({\bf 16_c}+{\bf 144}) t^5+({\bf 10_v}+{\bf 120})t^6-  (1+{\bf 45} )t^8
+{\bf 16_s} t^9\nn
\eea
 This reproduces the content of a massive spin-two  multiplet in $d=6,10$ dimensions.
 In particular, in $d=10$ one finds the $128_B-128_F$ degrees of freedom of
 the ten-dimensional open superstring.
 It would be nice to extend these results to higher string modes.

The pure spinor CFT's introduced here open new scenarios for studies of
holographic correspondences between gravity and minimal SYM gauge theories.
Indeed critical closed strings can be constructed in a similar way by tensoring
two copies of the CFT described here. In particular the spectra is given by the
square of (\ref{p0s},\ref{p1s},\ref{p16s}) and correspond to $N=2$ supergravities
in $d=4,6,10$. The non-perturbative spectrum of these theories
 always comprehends brane where the boundary gauge theories 
 described in the present work live. It would be nice to explore
applications of the pure spinor descriptions along these lines.

The techniques developed here also apply to a large class of interesting conformal
field theories defined via constraints. In particular in section (4)
we show how a critical string describing a two-dimensional CFT with ${\cal N}=(2,0)$
supersymmetry can be described in terms of two bosonic variables
satisfying a constraint. It would be nice to apply these ideas to the study of
elliptic genera of other constrained systems like strings moving on
algebraic surfaces.

 In \cite{Berkovits:2005hy}, the group structure of the pure spinor
space, which is a conical space (see \cite{gri-har}), was used
 to compute the zero-mode part of the partition function. One can ownder
 whether the same technique applies to massive states in terms of the associated
 Kac-Moody algebras.

\vskip 1cm

\section*{Acknowledgements}

We thank M. Bianchi, A. Lerda, Y. Oz, G. Policastro,  
H. Samtleben, R. Schiappa, E. Scheidegger, N. Wyllard  and in particular
R. Russo for discussions at various stages of this project.

\eject

\noindent {\bf \Large Appendices}
\begin{appendix}

\section{Massive Supermultiplets}
\label{secmult}

In this appendix we organize states in the string spectrum in multiplets
of minimal supersymmetry in $d$ dimensions.
Supermultiplets can be constructed
by acting with the raising supersymmetry charges on a highest weight states
(see \cite{Beisert:2003te} for details):
\be
{\bf [n_1,n_2...]}=
\sum_{ \epsilon_i=0,1 }\,(-t)^{2+\epsilon}\,
 {\rm dim}\,[n_1+\epsilon_i q_1^i,n_1+\epsilon_i q_2^i,..]^*
\label{mult0}
\ee
 with $\epsilon=\sum_i \epsilon_i$, $Q^i=[q^i_1,q^i_2,..]$ the weights of the raising supercharges
 $i=1,...S_d/2$ in the Dynkin basis. Finally
 $[n_1,n_2,..]^* \equiv [n_1,n_2,..]-[n_1-1,n_2,..] t^{-2}$ labels the
 $SO(d)$ Dynkin labels of
 the h.w.s.. The term with minus sign subtract
 the unphysical components, e.g. a massive vector in $d$ dimensions is
 written as $[100..]^*=[100..]-[000..]$ and
 so on. Alternatively one can write the polynomial with definitively positive
 coefficients in terms of $SO(d-1)$ representations. Here we prefer to
 keep $SO(d)$ covariance.

In table \ref{supercharges} we list the supercharges weights for minimal
supersymmetry in dimensions $d=4,6,10$.
\begin{table}[h]
\begin{center}
\begin{tabular}{|c|c|}
  \hline
 D & $Q^i$\\
\hline
  4 & $[-\ft12, 0], [-\ft12, 0]$ \\
  6 & $[0 ~0~ -1]_{\pm {1\over 2}} , [-1~ 0 ~1]_{\pm {1\over 2}}$ \\
  10 &
[0 0  0  0  -1],
[-1 0 1 0  -1],
[-1 1 -1 1 0],
[0-1  0  1  0 ],\\
 & [-1  1  0  -1  0 ],
 [0 -1  1  -1  0 ],
[0  0 -1  0  1 ],
[-1 0 0 0 1]\\
  \hline
\end{tabular}
\caption{Supercharges}
\label{supercharges}
\end{center}
\end{table}
Plugging the supercharges
into (\ref{mult0}) one finds that the content of vector multiplets
$-t^2{\bf [\ft12,\ft12]}$, $-t^2{\bf [1,0,0]_0}$, $-t^2{\bf [1,0,0,0,0]}$
match precisely
the polynomials (\ref{p0s}) describing the massless string
spectrum. In a similar way the content of a spin two multiplets
$t^4{\bf [1,1]}$, $t^4{\bf [2,0,0,0,0]}$ in $d=6,10$ match precisely
 the result $P^{\rm phys}_1(t)$ given by (\ref{p16s}) for the first
 massive string level in $d=6,10$.

\section{Ward Identities}

Here we derive eq.~(\ref{cic}) using anomalous Ward identities.  

The pure spinors are represented by a
Dirac spinor $\l^{A}$ ($A=1,\dots, 4$) satisfying  $\l \g^{m} \l =0$.
Therefore, there are eight
gauge invariant combinations
\be\label{xfourA}
J = :w_{A} \l^{A}:\,,~~~~~
J_{mn}= :w_{A} (\g_{mn})^{A}_{B} \l^{B}:\,, ~~~~~
J_{mnpq} = :w_{A} (\g_{mnpq})^{A}_{B} \l^{B}:\,.
\ee
The last one is a pseudoscalar quantity since it
can be rewritten in term of $\gamma^{5}$ as follows
$J_{mnpq} = \e_{mnpq} w_{A} (\g^{5})^{A}_{B} \l^{B}$ and we use the notation  
$w_{A} (\g^{5})^{A}_{B} \l^{B} \equiv J^{5}$. 

These gauge invariant operators satisfy the anomalous Ward identities
\bea\label{xfourC}
&& :J_{mn} \l^{B}: \gamma^{m}_{BC} - {1\over 2} :J \l^{B}: \gamma_{n, BC} = 
\alpha' \g_{n, BC} d \l^{C}\,,  \\
&& :J_{mn} \l^{A} \l^{B}: \gamma^{m}_{BC} - {1\over 2}:J \l^{A} \l^{B}:
\gamma_{n, BC} = {3 \a' \over 2} \l^{A} \p \l^{B} \g_{n, BC} + {\a' 
\over 2} \l^{D} \p \l^{B} (\gamma_{mn})_{D}^{~A} \gamma^{m}_{BC}\,,   \nonumber
\eea
 Now, in the d=4 case there is a new Ward identity relating the Lorentz generator  
 $J_{mn}$ and the pseudoscalar $J^{5}$. Notice that we 
 can rewrite the generator $J_{mn}$ as follows $J^{5}_{mn} = w_{A} 
 (\g^{5} \g^{mn})^{A}_{~B} \l^{B} = \e^{mnpq} J_{mn}$ where we have used the 
 relation $\gamma_{mn,AB}= \e_{mnpq} (\g^{5} \g^{pq})_{AB}$, thus we 
 have
\bea\label{xfourCA}
&& :J^{5}_{mn} \l^{B}: \gamma^{m}_{BC} - {1\over 2} :J^{5} \l^{B}: \gamma_{n, BC} = 
\alpha' (\g^{5}\g_{n})_{BC} \p \l^{C}\,,  \nonumber\\
&& :J^{5}_{mn} \l^{B} \l^{C}: \gamma^{m}_{CA} - {1\over 2}:J^{5} \l^{B} \l^{C}:
\gamma_{n, CA} = \\
&&~~~~~~~
\a' \l^{B}  (\g_{n}\g^{5} \p\l)_{A} + 
{\a' \over 2} (\g^{5}\l)^{B}  (\g_{n}\p\l)_{A} 
 - {\a' \over 2} (\gamma^{5}\gamma_{mn} \l)^{B}
 (\gamma^{m} \p \l)^{A} \,.  \nonumber
\eea
which are derived by using the Fiersz identitiy and the pure spinor 
condition.  Now, we can contract both sides of the second 
equation with $g^{n, AD}$ (and renaming $D \rightarrow A$) and we find 
\bea\label{xfourCB} 
&& - 
:J^{5}_{mn} \l^{B} \l^{C}: \gamma_{C}^{mn,A} - 2 :J^{5} \l^{B} \l^{A}: = \\
&&~~~~~~~
4 \a' \l^{B}  (\g^{5} \p\l)^{A} + 
2 {\a'} (\g^{5}\l)^{B}  (\g_{n}\p\l)^{A} 
 + {\a'\over 2} (\gamma^{5}\gamma_{mn} \l^{B})
 (\gamma^{mn} \p \l)^{A} \,.  \nonumber
\eea
and finally contracting both sides with $\g^{m}_{AB}$, 
we finally conclude that\footnote{The following conclusion 
can also be obtained using only the Ward identity (\ref{xfourC}), by 
first contracting with $\g^{n, AD}$ and then by $(\g^{5}\g^{m})_{AB}$.} 
\be\label{xfourCC}
\l \g^{5} \g^{m} \p \l =0\,.
\ee
Notice that the commuting nature of $\lambda$'s implies that 
$\l g^{5} \g^{m} \l =0$ and, the pure spinor condition implies that 
 $\l \g^{m} \p \l =0$, it turns out that due to pure spinor condition and 
 the properties of Dirac gamma matrices in d=4, we have that the 
 axial part of the bispinor $\l^{A} \p \l^{B}$ vanishes. Thus, this proves 
 eq.~(\ref{cic}) 
 
\end{appendix}



\begin{thebibliography}{99}

\bibitem{Berkovits:2000fe}
  N.~Berkovits,
  ``Super-Poincar\'e covariant quantization of the superstring,''
  JHEP {\bf 0004} (2000) 018
  [arXiv:hep-th/0001035].


\bibitem{Maldacena:1997re}
  J.~M.~Maldacena,
  Adv.\ Theor.\ Math.\ Phys.\  {\bf 2} (1998) 231
  [Int.\ J.\ Theor.\ Phys.\  {\bf 38} (1999) 1113]
  [arXiv:hep-th/9711200].




\bibitem{Berkovits:2002zk}
  N.~Berkovits,
  ``ICTP lectures on covariant quantization of the superstring,''
  arXiv:hep-th/0209059.


\bibitem{Berkovits:2000nn}
  N.~Berkovits,
  ``Cohomology in the pure spinor formalism for the superstring,''
  JHEP {\bf 0009} (2000) 046
  [arXiv:hep-th/0006003].



\bibitem{Berkovits:2004tw}
  N.~Berkovits and D.~Z.~Marchioro,
  ``Relating the Green-Schwarz and pure spinor formalisms for the
  superstring,''
  JHEP {\bf 0501}, 018 (2005)
  [arXiv:hep-th/0412198].



\bibitem{Aisaka:2004ga}
  Y.~Aisaka and Y.~Kazama,
  ``Relating Green-Schwarz and extended pure spinor formalisms by  similarity
  transformation,''
  JHEP {\bf 0404}, 070 (2004)
  [arXiv:hep-th/0404141].





\bibitem{Berkovits:2000ph}
  N.~Berkovits and B.~C.~Vallilo,
  ``Consistency of super-Poincare covariant superstring tree amplitudes,''
  JHEP {\bf 0007} (2000) 015
  [arXiv:hep-th/0004171].

\bibitem{Berkovits:2004px}
  N.~Berkovits,
  ``Multiloop amplitudes and vanishing theorems using the pure spinor formalism
  for the superstring,''
  JHEP {\bf 0409}, 047 (2004)
  [arXiv:hep-th/0406055].

\bibitem{Anguelova:2004pg}
  L.~Anguelova, P.~A.~Grassi and P.~Vanhove,
  ``Covariant one-loop amplitudes in D = 11,''
  Nucl.\ Phys.\ B {\bf 702}, 269 (2004)
  [arXiv:hep-th/0408171].



\bibitem{Grassi:2001ug}
  P.~A.~Grassi, G.~Policastro, M.~Porrati and P.~Van Nieuwenhuizen,
  ``Covariant quantization of superstrings without pure spinor constraints,''
  JHEP {\bf 0210} (2002) 054
  [arXiv:hep-th/0112162].

\bibitem{Grassi:2002tz}
  P.~A.~Grassi, G.~Policastro and P.~van Nieuwenhuizen,
  ``The massless spectrum of covariant superstrings,''
  JHEP {\bf 0211} (2002) 001
  [arXiv:hep-th/0202123].

\bibitem{Grassi:2002xv}
  P.~A.~Grassi, G.~Policastro and P.~van Nieuwenhuizen,
  ``On the BRST cohomology of superstrings with / without pure spinors,''
  Adv.\ Theor.\ Math.\ Phys.\  {\bf 7} (2003) 499
  [arXiv:hep-th/0206216].

\bibitem{Aisaka:2002sd}
  Y.~Aisaka and Y.~Kazama,
  ``A new first class algebra, homological perturbation and extension of pure
  spinor formalism for superstring,''
  JHEP {\bf 0302}, 017 (2003)
  [arXiv:hep-th/0212316].

\bibitem{Aisaka:2003mw}
  Y.~Aisaka and Y.~Kazama,
  ``Operator mapping between RNS and extended pure spinor formalisms for
  superstring,''
  JHEP {\bf 0308}, 047 (2003)
  [arXiv:hep-th/0305221].

\bibitem{Oda:2001zm}
  I.~Oda and M.~Tonin,
  ``On the Berkovits covariant quantization of GS superstring,''
  Phys.\ Lett.\ B {\bf 520}, 398 (2001)
  [arXiv:hep-th/0109051].

\bibitem{Matone:2002ft}
  M.~Matone, L.~Mazzucato, I.~Oda, D.~Sorokin and M.~Tonin,
  ``The superembedding origin of the Berkovits pure spinor covariant
  quantization of superstrings,''
  Nucl.\ Phys.\ B {\bf 639}, 182 (2002)
  [arXiv:hep-th/0206104].

\bibitem{Chesterman:2002ey}
  M.~Chesterman,
  ``Ghost constraints and the covariant quantization of the superparticle  in
  ten dimensions,''
  JHEP {\bf 0402}, 011 (2004)
  [arXiv:hep-th/0212261].





\bibitem{Berkovits:2005hy}
  N.~Berkovits and N.~Nekrasov,
  ``The character of pure spinors,''
  arXiv:hep-th/0503075.


\bibitem{Schiappa:2005mk}
  R.~Schiappa and N.~Wyllard,
  JHEP {\bf 0507}, 070 (2005)
  [arXiv:hep-th/0503123].


\bibitem{Bianchi:2003wx}
M.~Bianchi, J.~F. Morales, and H.~Samtleben, ``On stringy
{$AdS_5\times S^5$}
  and  holography,''
\href{http://www.arXiv.org/abs/hep-th/0305052}{{\tt
hep-th/0305052}}.


\bibitem{Beisert:2003te}
N.~Beisert, M.~Bianchi, J.~F. Morales, and H.~Samtleben, ``On the
spectrum of
  ads/cft beyond supergravity,'' {\em JHEP} {\bf 02} (2004) 001,
\href{http://www.arXiv.org/abs/hep-th/0310292}{{\tt
hep-th/0310292}}.


\bibitem{Beisert:2004di}
N.~Beisert, M.~Bianchi, J.~F. Morales, and H.~Samtleben,
``Higher
spin symmetry
  and {$\mathcal{N} = 4$} {SYM},'' {\em JHEP} {\bf 07} (2004) 058,
\href{http://www.arXiv.org/abs/hep-th/0405057}{{\tt
hep-th/0405057}}.

\bibitem{Morales:2004xc}
  J.~F.~Morales and H.~Samtleben,
  ``Higher spin holography for SYM in d dimensions,''
  Phys.\ Lett.\ B {\bf 607} (2005) 286
  [arXiv:hep-th/0411246].



\bibitem{Grassi:2005sb}
  P.~A.~Grassi and N.~Wyllard,
  ``Lower-dimensional pure-spinor superstrings,''
  arXiv:hep-th/0509140.

\bibitem{Wyllard:2005fh}
  N.~Wyllard,
  ``Pure-spinor superstrings in d = 2, 4, 6,''
  arXiv:hep-th/0509165.



\bibitem{Berkovits:2005bt}
  N.~Berkovits,
  ``Pure spinor formalism as an N = 2 topological string,''
  arXiv:hep-th/0509120.

\bibitem{Chandia:2005fi}
  O.~Chandia,
  ``D = 4 pure spinor superstring and N = 2 strings,''
  arXiv:hep-th/0509185.

\bibitem{Movshev:2005cq}
  M.~Movshev,
  ``Yang-Mills theories in dimensions 3,4,6,10 and Bar-duality,''
  arXiv:hep-th/0503165.



\bibitem{Klebanov:2004ya}
  I.~R.~Klebanov and J.~M.~Maldacena,
  Int.\ J.\ Mod.\ Phys.\ A {\bf 19} (2004) 5003
  [arXiv:hep-th/0409133].



\bibitem{Berkovits:2002qx}
  N.~Berkovits and O.~Chandia,
  ``Massive superstring vertex operator in D = 10 superspace,''
  JHEP {\bf 0208} (2002) 040
  [arXiv:hep-th/0204121].


\bibitem{DiFrancesco:1997nk}
  P.~Di Francesco, P.~Mathieu and D.~Senechal,
  ``Conformal field theory,'', 1997 Spinger-Verlag, New York.


\bibitem{gri-har}
P.~Griffiths, J.~Harris,
``Principles of algebraic geometry,'' 1978 - New York: Wiley

\bibitem{Grassi:2004tv}
  P.~A.~Grassi and G.~Policastro,
  ``Super-Chern-Simons theory as superstring theory,''
  arXiv:hep-th/0412272.

\end{thebibliography}
\end{document}